\documentclass{article} %
\usepackage{colm2024_conference}

\usepackage{booktabs}
\usepackage{graphicx}
\usepackage{enumitem}
\usepackage{wrapfig}
\usepackage{algorithm}
\usepackage{algpseudocode}
\usepackage{multicol}

\usepackage{CJKutf8}

\usepackage{microtype}
\usepackage{amsmath}
\usepackage{amsfonts}
\usepackage{colortbl}
\usepackage[utf8]{inputenc}
\usepackage[T1]{fontenc}
\definecolor{lightgray}{rgb}{0.9,0.9,0.9}
\usepackage{caption}
\usepackage{subcaption}
\usepackage{xcolor}
\usepackage{setspace}
\usepackage{url}
\usepackage{multirow}
\usepackage{colortbl}
\usepackage{tabularx}
\usepackage{xltabular}
\usepackage{blindtext}
\usepackage{pgfplots}
\pgfplotsset{compat=1.18} 
\usepackage{tikz}
\usetikzlibrary{er,positioning,bayesnet,arrows.meta,fit,backgrounds,decorations.pathreplacing,shadows.blur,calc}
\usepackage{makecell}
\usepackage{tipa}
\usepackage{siunitx}
\usepackage{nicefrac}
\usepackage{tocloft}
\usepackage{listings}
\usepackage[raster,skins]{tcolorbox} %
\usepackage{xltabular}
\usepackage{adjustbox}
\usepackage{xurl}
\usepackage{multirow}
\usepackage{threeparttable}
\usepackage{xcolor}

\usepackage{amsmath,amsfonts,bm}

\def\eqref#1{equation~\ref{#1}}

\def\1{\bm{1}}

\DeclareMathAlphabet{\mathsfit}{\encodingdefault}{\sfdefault}{m}{sl}
\SetMathAlphabet{\mathsfit}{bold}{\encodingdefault}{\sfdefault}{bx}{n}

\title{Qwen-Audio-VAE Technical Report}

\author{
\bf Qwen Team
}

\newcommand{\method}{Qwen-Audio-VAE\xspace}

\begin{document}

\maketitle

\begin{abstract}
We introduce \textbf{Qwen-Audio-VAE}, a suite of low-bitrate, fast-encoding continuous audio autoencoders designed for scalable general audio generation. The model is built around a simple but important principle: an audio VAE should not only reconstruct diverse audio with high fidelity, but also produce compact latent representations fast enough to support large-scale text-to-audio training. Qwen-Audio-VAE combines a causal encoder-decoder, window Transformer blocks, and multi-discriminator training to achieve a strong balance between reconstruction quality and compression rate. The model is trained at scale on 5 million hours of multi-domain audio, enabling robust reconstruction across heterogeneous acoustic conditions. To further improve computational efficiency, we adopt an asymmetric encoder-decoder backbone and introduce latency-aware encoder pruning to maximize encoding throughput. Experiments on public speech, music, and sound reconstruction benchmarks show that Qwen-Audio-VAE generalizes well across diverse audio domains and is particularly efficient, requiring only 541 ms to encode 32 minutes of audio. Overall, Qwen-Audio-VAE provides a high-quality, compact, and high-throughput representation backbone for efficient general audio generation.

\end{abstract}

\begin{figure}[tbh]
    \centering
    \includegraphics[width=0.9\textwidth]{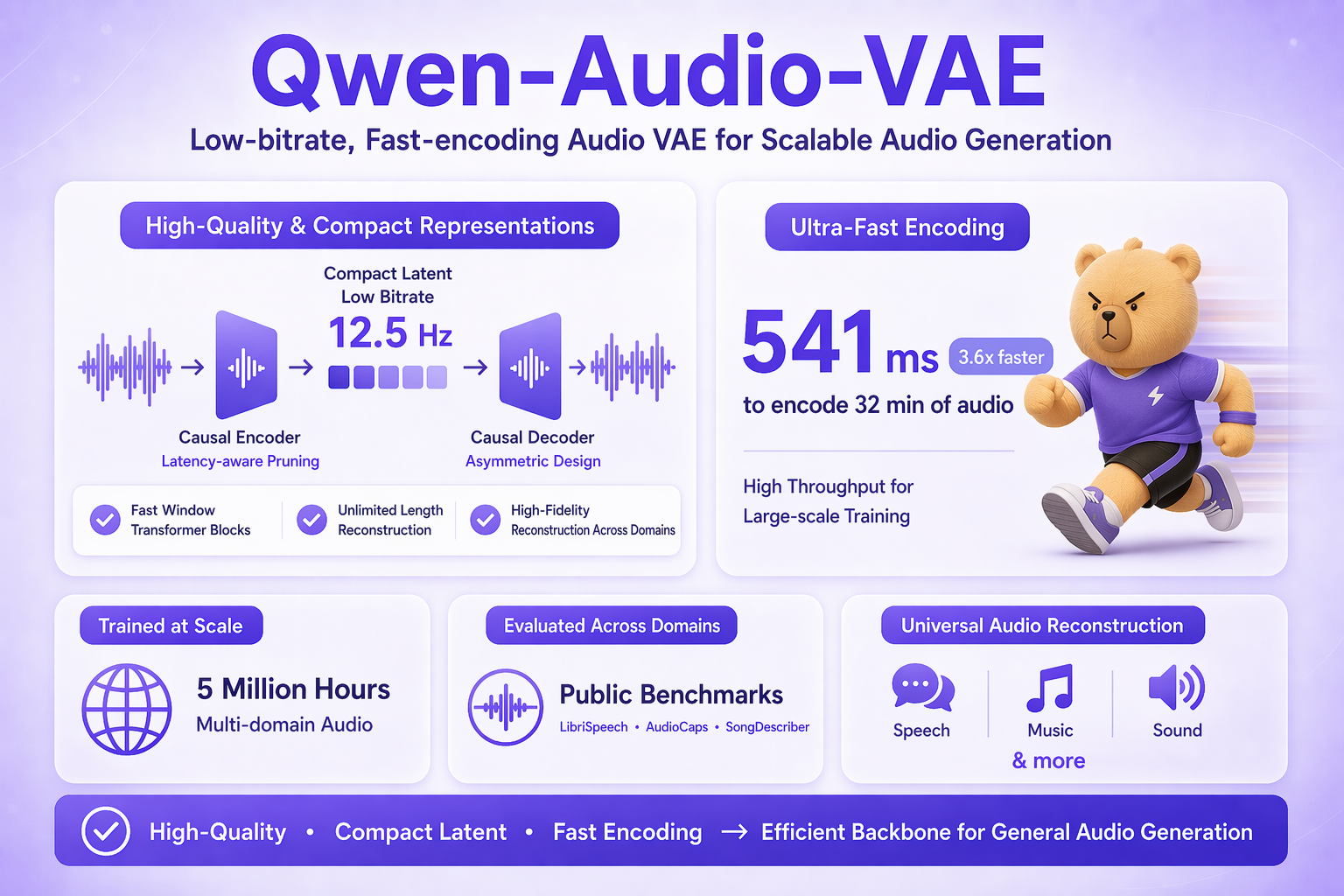}
    \caption{\textbf{\method}: a low-bitrate, fast-encoding continuous audio
    autoencoder with high-fidelity reconstruction across speech, music,
    and sound and fast encoding (\textbf{541\,ms}\, for 32 min of audio), making it an efficient backbone for general audio generation.}
    \label{fig:intro}
\end{figure}

\section{Introduction}
\label{sec:intro}

Multimodal foundation models~\citep{openai2024gpt4-o,gemini2.5,qwen35omni} have made text-to-audio-video generation~\citep{yang2025cogvideox,chen2025skyreels,wan2025wan,seedance2026seedance} a fast-growing direction. These systems~\citep{vqgan,dit} do not model waveforms directly but generate in the compact latent space of an autoencoder~\citep{audioldm,stableaudioopen,mmaudio,hunyuanfoley,klingfoley}. The autoencoder is thus foundational: it determines the representation the generator must learn and caps the audio quality achievable after decoding.

For \emph{general} audio generation, a good autoencoder must satisfy three requirements at once. \emph{First, high-fidelity reconstruction} across diverse audio, since reconstruction error upper-bounds the quality the generator can ultimately produce. \emph{Second, a compact latent}, because a shorter latent sequence directly reduces the training cost of downstream diffusion transformers~\citep{dit}. \emph{Third, high-throughput encoding}: in large-scale text-to-audio training, latents are often computed online, so a slow encoder becomes a training bottleneck that limits how much audio each step can process.

Existing representations, however, tend to prioritize a subset of these requirements rather than balancing all three together. Discrete audio codecs~\citep{SoundStream,Encodec,dac,wavtokenizer,SpeechTokenizer} compress speech very well, but their quantization error tends to weaken reconstruction on noisy general sound and higher-fidelity music. Continuous audio VAEs~\citep{vae} avoid this quantization loss and pair naturally with diffusion- and flow-based generators, making them attractive for high-fidelity general audio. Their limitations lie instead in efficiency and coverage: most run at relatively high frame rates~\citep{mmaudio,hunyuanfoley,stableaudioopen}, which lengthen the latent sequence and increase downstream DiT cost, and many~\citep{peng2025vibevoice} are designed for a single domain such as clean speech rather than the heterogeneous in-the-wild audio that general audio generation demands.

To close this gap, we present \textbf{Qwen-Audio-VAE}, a 12.5\,Hz continuous audio autoencoder that targets all three requirements together. Our main contributions are:

\begin{itemize}
\item We train Qwen-Audio-VAE on \textbf{5 million hours} of multi-domain audio, combining a causal encoder-decoder, window Transformer blocks, and multi-discriminator training to obtain a compact, high-fidelity, and domain-robust representation.
\item We improve encoding efficiency through an asymmetric encoder-decoder backbone, latency-aware encoder pruning, and concentrating the heaviest window Transformer at the lowest frame rate, reducing 64$\times$30s encoding latency from \textbf{1957 ms to 541 ms} (a 3.62$\times$ speedup) without degrading reconstruction.
\item We show that Qwen-Audio-VAE generalizes well across public speech, sound, and music benchmarks (LibriSpeech, AudioCaps, SongDescriber), while also improving downstream text-to-audio generation quality.
\end{itemize}

\section{Architecture}

\begin{figure}[tbh]
    \centering
    \definecolor{cEnc}{HTML}{4B3B96}
    \definecolor{cWT}{HTML}{6E4AA6}
    \definecolor{cVAE}{HTML}{9B3FA6}
    \definecolor{cDisc}{HTML}{5B6472}
    \newcommand{\shadowrect}[4]{\fill[black!7, rounded corners=3pt]
        ({#1-#3/2+0.05},{#2-#4/2-0.07}) rectangle ({#1+#3/2+0.05},{#2+#4/2-0.07});}
    \resizebox{\textwidth}{!}{%
    \begin{tikzpicture}[
        font=\small,
        >={Stealth[length=2.4mm]},
        wt/.style={draw=cWT!80, line width=0.8pt, rounded corners=3pt, fill=cWT!10,
            minimum width=2.0cm, minimum height=1.55cm, align=center},
        lat/.style={draw=cVAE!80, line width=0.8pt, rounded corners=3pt, fill=cVAE!8,
            minimum width=1.4cm, minimum height=1.75cm},
        disc/.style={draw=cDisc!80, line width=0.7pt, rounded corners=3pt, fill=cDisc!7,
            minimum width=3.6cm, minimum height=0.9cm, align=center},
        io/.style={align=center, font=\small\itshape, text=black!70},
        flow/.style={->, line width=1pt, draw=black!72},
        rate/.style={font=\footnotesize, text=black!55},
    ]
    \draw[cEnc!85, line width=0.5pt, samples=100, domain=-0.45:0.45, smooth, variable=\x]
        plot ({0.35+\x*1.2},{0.42*sin(\x*2600)*exp(-\x*\x*9)});
    \node[rate] at (0.35,1.0) {24\,kHz};
    \node[io] at (0.35,-0.8) {input\\waveform};
    \fill[black!7, rounded corners=2pt] (1.55,1.13)--(3.55,0.68)--(3.55,-0.82)--(1.55,-1.27)--cycle;
    \draw[draw=cEnc!80, line width=0.8pt, fill=cEnc!14, rounded corners=2pt]
        (1.5,1.2)--(3.5,0.75)--(3.5,-0.75)--(1.5,-1.2)--cycle;
    \node[align=center, font=\small\bfseries, text=black!85] at (2.5,0) {Conv\\Encoder};
    \shadowrect{5.2}{0}{2.0}{1.55}
    \node[wt] (pre) at (5.2,0) {\textbf{Window}\\\textbf{Transformer}\\{\footnotesize\itshape (pre)}};
    \shadowrect{7.35}{0}{1.4}{1.75}
    \node[lat] (vae) at (7.35,0) {};
    \node[font=\small\bfseries, text=black!85] at (7.35,0.55) {Latent $z$};
    \draw[cVAE!55, line width=0.5pt] (6.85,-0.28)--(7.85,-0.28);
    \draw[cVAE!85, line width=0.7pt, samples=50, domain=-0.5:0.5, smooth, variable=\x]
        plot ({7.35+\x*0.72},{-0.28+0.45*exp(-\x*\x*7)});
    \draw[cVAE!45, line width=0.4pt] (7.35,-0.28)--(7.35,0.16);
    \fill[cVAE!85] ({7.35+0.22},{-0.28+0.45*exp(-0.22*0.22*7)}) circle (1.2pt);
    \node[rate] at (7.35,-0.62) {128-d};
    \shadowrect{9.5}{0}{2.0}{1.55}
    \node[wt] (post) at (9.5,0) {\textbf{Window}\\\textbf{Transformer}\\{\footnotesize\itshape (post)}};
    \fill[black!7, rounded corners=2pt] (11.05,0.78)--(13.45,1.33)--(13.45,-1.47)--(11.05,-0.92)--cycle;
    \draw[draw=cEnc!80, line width=0.8pt, fill=cEnc!16, rounded corners=2pt]
        (11.0,0.85)--(13.4,1.4)--(13.4,-1.4)--(11.0,-0.85)--cycle;
    \node[align=center, font=\small\bfseries, text=black!85] at (12.2,0) {Conv\\Decoder};
    \draw[cEnc!85, line width=0.5pt, samples=100, domain=-0.45:0.45, smooth, variable=\x]
        plot ({14.45+\x*1.2},{0.42*sin(\x*2600)*exp(-\x*\x*9)});
    \node[rate] at (14.45,1.0) {24\,kHz};
    \node[io] at (14.45,-0.8) {reconstructed\\waveform};
    \draw[flow] (0.95,0)--(1.5,0);
    \draw[flow] (3.5,0)--(pre.west) node[midway,above=1pt,rate]{$\downarrow 4$};
    \draw[flow] (pre.east)--(vae.west);
    \draw[flow] (vae.east)--(post.west);
    \draw[flow] (post.east)--(11.0,0) node[midway,above=1pt,rate]{$\uparrow 4$};
    \draw[flow] (13.4,0)--(13.9,0);
    \node[rate] at (2.9,1.45) {50\,Hz};
    \node[rate] at (12.0,1.62) {50\,Hz};
    \draw[decorate, decoration={brace, amplitude=5pt, mirror}, draw=black!45]
        ([yshift=-9pt]pre.south west) -- ([yshift=-9pt]post.south east)
        node[midway, below=5pt, rate]{12.5\,Hz latent space};
    \shadowrect{5.3}{-3.7}{3.6}{0.9}
    \shadowrect{9.5}{-3.7}{3.6}{0.9}
    \shadowrect{5.3}{-4.8}{3.6}{0.9}
    \shadowrect{9.5}{-4.8}{3.6}{0.9}
    \node[disc] (d1) at (5.3,-3.7)  {Multi-Period};
    \node[disc] (d2) at (9.5,-3.7)  {Multi-Resolution STFT};
    \node[disc] (d3) at (5.3,-4.8)  {Multi-Scale STFT};
    \node[disc] (d4) at (9.5,-4.8)  {Sub-band CQT};
    \begin{scope}[on background layer]
      \node[draw=black!35, dashed, rounded corners=4pt, fill=cDisc!4,
            fit=(d1)(d2)(d3)(d4), inner sep=10pt] (dbox) {};
    \end{scope}
    \node[rate, above=2pt of dbox] {Discriminators\,(training only)};
    \coordinate (obl) at (14.45,-1.15);
    \draw[flow] (obl) |- (dbox.east);
    \end{tikzpicture}%
    }
    \caption{Overview of \method. A causal convolutional encoder downsamples the
    waveform to 50\,Hz; a bottleneck reduces it to a \textbf{12.5\,Hz}
    diagonal-Gaussian latent between two window Transformers; and an asymmetric
    decoder reconstructs the waveform. During training, four discriminators
    provide adversarial supervision.}
    \label{fig:ovewview_arch}
\end{figure}
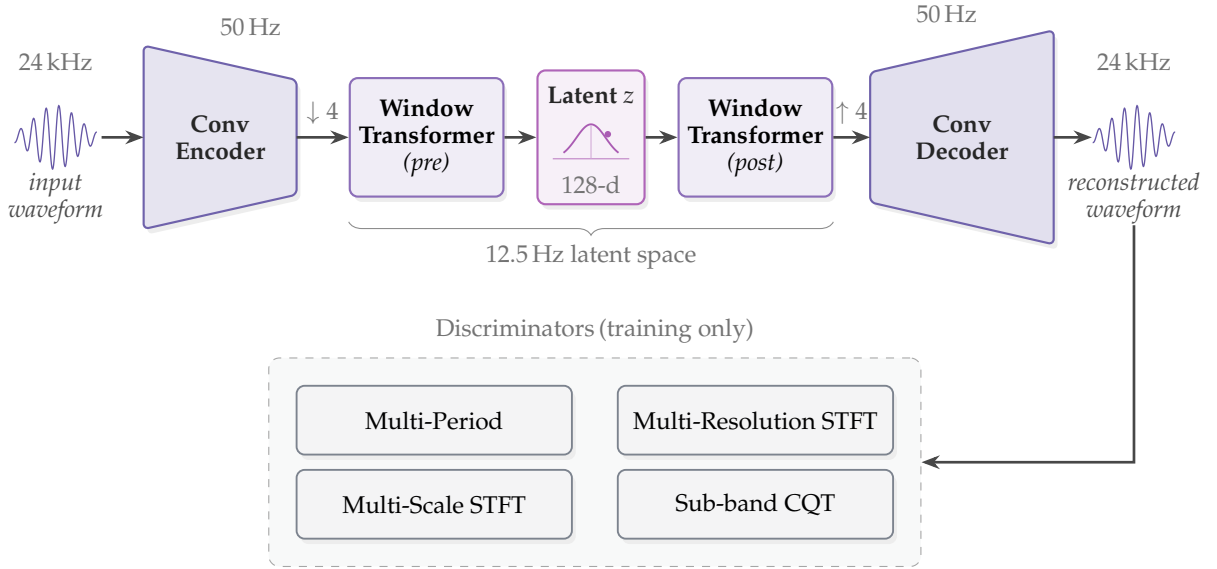

\subsection{Overview}

Qwen-Audio-VAE is a continuous variational autoencoder for 24\,kHz mono audio, designed as a representation backbone for scalable general audio generation rather than a standalone codec. As illustrated in Figure~\ref{fig:ovewview_arch}, a causal encoder $E$ maps a waveform $x\in\mathbb{R}^{T}$ to the parameters of a diagonal-Gaussian posterior over a latent sequence $z\in\mathbb{R}^{D\times L}$, from which a decoder $G$ reconstructs the signal as $\hat{x}=G(z)$; the two are trained end-to-end with the spectral-reconstruction and adversarial objectives of Section~\ref{sec:losses}. Architecturally, the model pursues the three requirements set out in Section~\ref{sec:intro}: high-fidelity reconstruction, a compact latent, and high-throughput encoding---and the rest of this section is organized around how each is achieved.

The compact latent is the design's central mechanism, so we quantify it first. The encoder downsamples the waveform by $480\times$ to 50\,Hz, and the bottleneck compresses it by a further $4\times$, for a total $1920\times$ reduction to a \textbf{12.5\,Hz} latent. A 30\,s clip therefore maps to only about $375$ latent frames, versus roughly $600$--$2250$ at the 20--75\,Hz rates common in prior VAEs. Because the self-attention cost of a diffusion-transformer (DiT) generator~\citep{dit} grows as $\mathcal{O}(L^2)$ in the latent length $L$, this low frame rate is the primary lever on downstream training cost. The other two requirements are then met around this compact latent: the training objectives (Section~\ref{sec:losses}) preserve fine acoustic detail despite the aggressive compression, while an asymmetric encoder--decoder and latency-aware encoder pruning (Section~\ref{sec:encoder-accel}) keep the latents themselves cheap to produce.

\subsection{Model Components}
\label{sec:components}

Qwen-Audio-VAE follows a VAE-GAN design: a \emph{generator}---a causal encoder--decoder with a continuous latent bottleneck---that maps a waveform to a compact latent and reconstructs it, and a bank of \emph{discriminators} that provide adversarial supervision during training. The full configuration is given in Table~\ref{tab:model-config} (Appendix~\ref{sec:appendix-config}).

\paragraph{Generator.} The generator is a causal convolutional encoder--decoder with a continuous latent bottleneck, comprising four modules in signal-flow order.
\begin{enumerate}
\item \textbf{Causal encoder.} A DAC-style~\citep{dac} causal convolutional stack downsamples the waveform to 50\,Hz. An initial convolution lifts the mono input to 24 channels, and four strided residual blocks with strides $(8,5,4,3)$ reduce the temporal resolution by $480\times$ while widening the channels $24\!\to\!128\!\to\!256\!\to\!512\!\to\!1024$. Each block pairs a Snake-activated residual unit with a strided convolution, and causal padding throughout preserves streaming and chunk-wise inference. The encoder outputs a feature map $h\in\mathbb{R}^{1024\times T/480}$.
\item \textbf{Continuous latent bottleneck.} In place of vector quantization, we use a continuous diagonal-Gaussian latent~\citep{vae}. A further $4\times$ temporal downsampling brings $h$ to 12.5\,Hz, and a linear projection produces the per-frame posterior mean and log-variance $\mu,\log\sigma^2\in\mathbb{R}^{D\times L}$ with $D=128$; latents are then sampled with the reparameterization trick
\begin{equation}
z = \mu + \sigma\odot\epsilon,\qquad \epsilon\sim\mathcal{N}(\mathbf{0},\mathbf{I}),
\label{eq:reparam}
\end{equation}
which keeps sampling differentiable. The continuous formulation avoids quantization error and matches the continuous trajectories modeled by diffusion- and flow-based generators.
\item \textbf{Window Transformer blocks.} Two windowed Transformers (8 layers, width 1024, 16 heads, attention window 72) flank the latent projection, both operating at 12.5\,Hz. They supply the long-range context and modeling capacity that convolutions alone lack under such aggressive compression. Placing them at the lowest frame rate is deliberate: as the heaviest modules in the model, running them on the shortest sequence minimizes their cost and keeps them off the encoder's high-resolution critical path.
\item \textbf{Asymmetric decoder.} A transposed-convolutional decoder inverts the $1920\times$ compression back to 24\,kHz. Because decoding lies off the critical path of large-scale latent extraction (Section~\ref{sec:experiment}), we give it substantially more capacity than the encoder---base width 1536 and three residual units per block with dilations $\{1,3,9\}$---which is the source of the encoder--decoder asymmetry.
\end{enumerate}

\paragraph{Discriminators.} During training, the reconstruction $\hat{x}$ is critiqued against the reference $x$ by a bank of $K$ discriminators $\{D_k\}_{k=1}^{K}$, each specialized to a different acoustic aspect of the signal. No single representation exposes every kind of artifact, so we combine complementary time- and frequency-domain critics; together they drive the reconstruction toward realism across speech, music, and general sound.
\begin{enumerate}
\item \textbf{Multi-period discriminator}~\citep{hifigan}, which reshapes the waveform along periods $\{2,3,5,7,11\}$ to capture periodic structure such as pitch and voicing in speech and singing.
\item \textbf{Multi-resolution STFT discriminator}~\citep{dac}, which splits the magnitude spectrum into sub-bands at several FFT sizes to enforce fidelity across frequency bands, especially the high frequencies.
\item \textbf{Multi-scale STFT discriminator}~\citep{Encodec}, which operates on complex spectrograms at multiple resolutions to capture both fine transients and coarse time--frequency structure.
\item \textbf{Sub-band CQT discriminator}, whose constant-$Q$, log-frequency resolution matches the harmonic spacing of music and singing, sharpening tonal and harmonic detail.
\end{enumerate}
The exact periods, FFT sizes, and CQT resolutions are listed in Appendix~\ref{sec:appendix-disc}. The discriminators enter only the adversarial objective (Section~\ref{sec:losses}) and add no inference-time cost.

\subsection{Training Objective}
\label{sec:losses}

High-fidelity reconstruction requires supervising both the global spectral envelope and the fine detail that $\ell_1$ spectral losses alone tend to over-smooth. We therefore combine multi-scale spectral reconstruction losses, which provide a stable and perceptually meaningful gradient, with the adversarial discriminators of Section~\ref{sec:components}, which sharpen fine structure and restore realistic high-frequency content. Let $x$ denote the reference and $\hat{x}=G(z)$ the reconstruction; the generator minimizes a weighted sum of spectral, adversarial, feature-matching, and KL terms,
\begin{equation}
\mathcal{L}_{G} = \lambda_{\text{mel}}\mathcal{L}_{\text{mel}} + \lambda_{\text{stft}}\mathcal{L}_{\text{stft}} + \lambda_{\text{adv}}\mathcal{L}_{\text{adv}} + \lambda_{\text{fm}}\mathcal{L}_{\text{fm}} + \lambda_{\text{kl}}\mathcal{L}_{\text{kl}}.
\label{eq:gen}
\end{equation}
The two spectral terms, $\mathcal{L}_{\text{mel}}$ and $\mathcal{L}_{\text{stft}}$, are multi-scale $\ell_1$ distances between the log-mel and magnitude-STFT representations of $x$ and $\hat{x}$, computed over a shared set of analysis windows (seven scales from 32 to 2048 samples) so that the coarse envelope and fine harmonic detail are supervised together; for the STFT term we use the perceptually weighted \texttt{auraloss} \texttt{MultiResolutionSTFTLoss}~\citep{auraloss}.

The adversarial and feature-matching terms are computed over the discriminator bank of Section~\ref{sec:components}. Adversarial training follows the least-squares GAN form: the generator drives each $D_k(\hat{x})$ toward $1$, while every discriminator $D_k$ is trained to separate $x$ from $\hat{x}$, averaged over the $K$ discriminators. A feature-matching term $\mathcal{L}_{\text{fm}}$ complements this by averaging the $\ell_1$ distance between the intermediate feature maps that each $D_k$ extracts from $x$ and $\hat{x}$; acting as a perceptual loss in discriminator space, it is the main stabilizer of adversarial training. Finally, $\mathcal{L}_{\text{kl}}$ is the standard closed-form KL divergence that regularizes the diagonal-Gaussian posterior toward the prior $\mathcal{N}(\mathbf{0},\mathbf{I})$. In the final configuration, the spectral objective is weighted toward the STFT term ($\lambda_{\text{stft}}=20$, alongside a multi-scale mel term); the adversarial and feature-matching terms carry unit weight ($\lambda_{\text{adv}}=\lambda_{\text{fm}}=1$); and the KL weight is kept deliberately small ($\lambda_{\text{kl}}=10^{-6}$), so that reconstruction quality remains the dominant objective while the latent stays only lightly regularized.

\begin{table}[htbp]
\centering
\caption{Reconstruction collapse when naively reducing the entire encoder channel size, motivating the first-layer-only reduction.}
\label{tab:naive-channel-cut}
\small
\begin{tabular}{lcccc}
\toprule
Variant & PESQ wb/nb ↑ & STOI ↑ & MR-STFT ↓ & SI-SDR ↑ \\
\midrule
Before (full channels) & \textbf{3.09 / 3.44} & \textbf{0.87} & \textbf{0.848} & \textbf{8.34} \\
After (all channels reduced) & 1.91 / 2.37 & 0.74 & 1.08 & -2.80 \\
\bottomrule
\end{tabular}
\end{table}

\begin{table}[htbp]
\centering
\caption{Per-layer encoder time profile (ms). The first layer dominates because it operates at the highest temporal resolution.}
\label{tab:layer-profile}
\small
\begin{tabular}{lc}
\toprule
Encoder layer & Time (ms) \\
\midrule
Layer 1 (highest resolution) & 305.8 \\
Layer 2 & 78.1 \\
Layer 3 & 36.4 \\
Layer 4 & 197.4 \\
\bottomrule
\end{tabular}
\end{table}

\subsection{Encoder Acceleration}
\label{sec:encoder-accel}

In text-to-audio integration, encoder latency---not the decoder---was the throughput bottleneck: encoding 64 clips of 30\,s initially took $\approx$1957\,ms. We therefore treat the encoder as a system-critical module and optimize its latency while holding the reconstruction quality and the 12.5\,Hz latent rate fixed.

The guiding principle is that computation at high temporal resolution is disproportionately expensive: early layers process long sequences, so even moderate channel counts dominate latency, whereas later layers run at lower resolution and are cheaper places to spend capacity. This is already reflected in the architecture---the window Transformers, the model's largest blocks, operate at the 12.5\,Hz bottleneck rather than at high resolution---and we push it further with three encoder-specific steps.

\textbf{Step 1: stride re-allocation.} We re-balance the strides so that most parameters act on lower-resolution intermediate features, cutting latency from \textbf{1957\,ms to 1361\,ms} with negligible quality loss.

\textbf{Step 2: residual-unit pruning.} Dilated residual units add convolutional cost, so we keep a single dilation-1 residual unit per encoder block. Latency drops from \textbf{1361\,ms to 747\,ms}.

\textbf{Step 3: first-layer channel reduction.} Layer-wise profiling attributes most cost to the first layer, which runs at the highest resolution (Table~\ref{tab:layer-profile}). Shrinking \emph{all} channels collapses quality---PESQ (wb/nb) falls from 3.09/3.44 to 1.91/2.37 and SI-SDR from 8.34 to $-2.80$ (Table~\ref{tab:naive-channel-cut})---so we reduce only the first-layer width from \textbf{64 to 24}, preserving later-layer capacity. Latency falls from \textbf{747\,ms to 541\,ms} without harming convergence.

The three steps compound to a $3.62\times$ speedup ($1957\!\to\!541$\,ms) at almost unchanged reconstruction quality, making fast encoding a core practical advantage: it shortens offline latent extraction and enables larger text-to-audio batches at the same step time. Section~\ref{sec:experiment} quantifies the end-to-end speedup and its downstream impact.

\section{Data}
\label{sec:data}

Training a domain-general audio VAE at scale poses three data challenges: assembling a corpus diverse enough to avoid overfitting to any single acoustic condition, curating it reliably so that corrupted or mislabeled audio does not degrade reconstruction, and feeding it efficiently enough to keep large GPU clusters saturated. We address these in turn.

\subsection{Data Collection}

Qwen-Audio-VAE is trained on \textbf{5 million hours} of multi-domain audio to support robust reconstruction across heterogeneous acoustic conditions. The corpus spans three major categories (Table~\ref{tab:data-composition}):

\begin{itemize}
\item \textbf{Speech} ($\sim$3M h): English and Chinese speech, multilingual dialects, historical ASR data, and podcasts.
\item \textbf{Music} ($\sim$1.3M h): instrumental and vocal music.
\item \textbf{Sound} ($\sim$0.8M h): online sound effects, natural and environmental sound, and audio from online videos.
\end{itemize}

This mixture matters for general audio generation because each category stresses a different aspect of reconstruction. Speech emphasizes intelligibility, speaker characteristics, prosody, and periodic structure. Music emphasizes harmonic structure, timbre, rhythm, and wide-band frequency detail. Sound and online-video audio introduce transient events, noisy environments, overlapping sources, and highly diverse acoustic scenes. Training on all three together makes the continuous latent more robust and reduces the risk of overfitting to clean speech or narrow acoustic conditions. All audio is resampled to \textbf{24\,kHz} mono before training.

\begin{table}[htbp]
\centering
\caption{Composition of the Qwen-Audio-VAE training corpus.}
\label{tab:data-composition}
\small
\begin{tabular}{llc}
\toprule
Domain & Data Categories & Hours \\
\midrule
Speech & EN/ZH, dialects, ASR, podcasts & $\sim$3M \\
Music & Instrumental, vocal & $\sim$1.3M \\
Sound & Online SFX, natural sound, sound from online video & $\sim$0.8M \\
\midrule
Total & Mixed & $\sim$5M \\
\bottomrule
\end{tabular}
\end{table}

\subsection{Data Curation and Quality Control}
\label{sec:data-curation}

\begin{figure}[htbp]
\centering
\includegraphics[width=\textwidth]{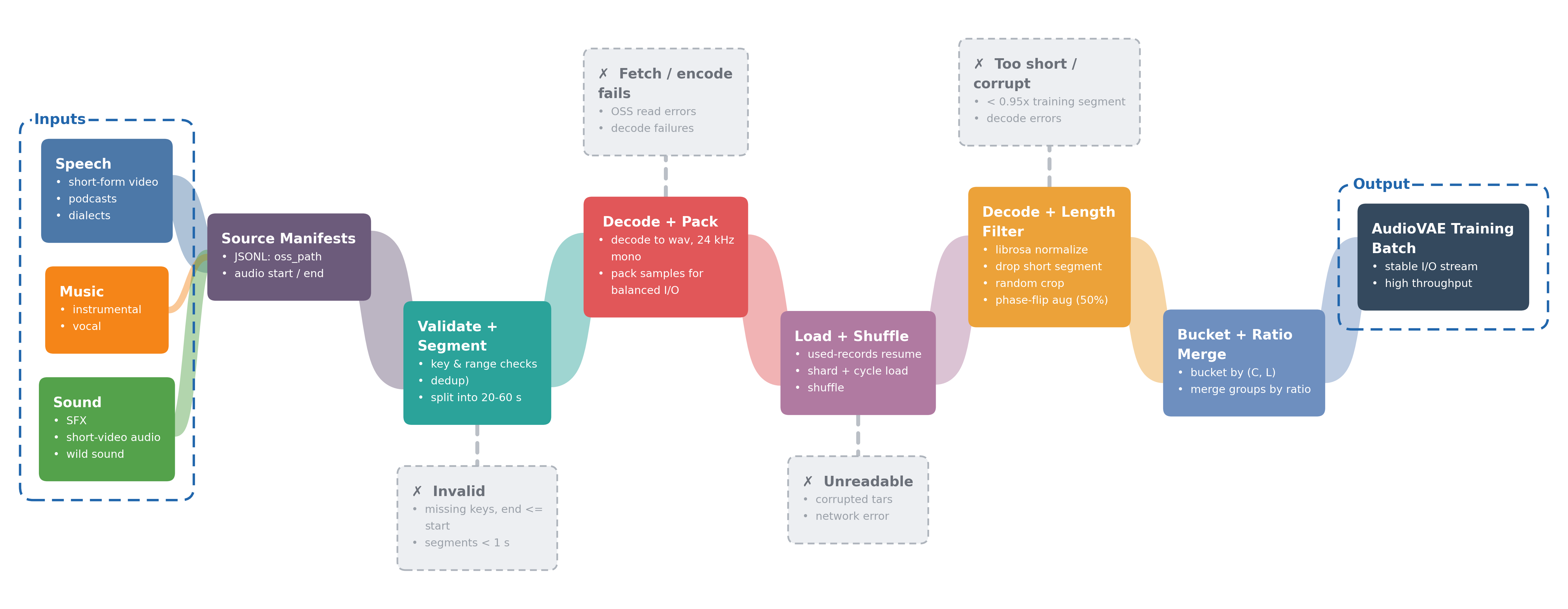}
\caption{Qwen-Audio-VAE data pipeline. The multi-domain raw audio pool flows through accessibility, duration, decode/sample-rate, and content-quality stages; abnormal samples are skipped, repacked, filtered, or down-weighted rather than silently entering training. Retained samples are grouped into tar shards with offset-based reading, improving training throughput from 0.90 to 1.07 it/s.}
\label{fig:data-pipeline}
\end{figure}

Reconstruction training is sensitive to corrupted audio, incorrect metadata, and distribution imbalance, so the raw multi-domain pool is curated before it is packed for training. As illustrated in Figure~\ref{fig:data-pipeline}, every sample passes through a sequence of validation stages, and abnormal samples are handled explicitly at each stage rather than being allowed to silently enter training.
\begin{itemize}
\item \textbf{Accessibility check}: Missing and unrecoverable samples are \emph{skipped}.
\item \textbf{Duration filtering}: duration-mismatch checks, including a fix for a previously buggy ${>}60$\,s filter; recoverable and false-positive cases are \emph{repacked}.
\item \textbf{Decode / sample-rate check}: corrupted audio and sample-rate mismatches are \emph{filtered} out.
\item \textbf{Content-quality stratification}: per-domain cleanliness and class-consistency checks; low-quality or domain-inconsistent samples---e.g., vocals leaking into instrumental music---are \emph{down-weighted or excluded}.
\end{itemize}
Beyond these validity checks, the pipeline supports multi-level filtering and data stratification: speech, music, and sound are weighted-sampled separately, and problematic subsets can be traced through their identifiers. This is useful for diagnosing domain-specific artifacts, such as high-frequency loss in music, transient smearing in sound effects, or robustness in noisy speech. Because Qwen-Audio-VAE serves as a representation backbone for text-to-audio training, reliable metadata and stable sample scheduling matter as much as raw data scale.

\subsection{Data Packaging and Loading}
\label{sec:data-packaging}

At the five-million-hour scale, training is limited not only by model computation but also by data I/O: raw collections contain many small files with heterogeneous metadata, which makes random access expensive and unstable. We therefore pack the curated samples into binary tar shards with a WebDataset-style index, storing each sample together with metadata such as offset, duration, sample rate, and identifier. A tar reader can then seek directly to a sample's offset and read it into memory, avoiding random small-file IO, while a \emph{history-records} module tracks which samples have been consumed---enabling resume from interruptions and stable, deduplicated scheduling across long-running jobs.

In practice, this strategy increases training throughput by approximately \textbf{18\%}. Although it leaves the model unchanged, it materially accelerates experiment iteration and makes five-million-hour-scale training more practical.

\section{Evaluation}
\label{sec:experiment}

We now examine whether Qwen-Audio-VAE meets its three design goals---high-fidelity reconstruction, a compact latent, and fast encoding---and whether these properties carry over to downstream text-to-audio training.

\subsection{Evaluation Setup}

We evaluate Qwen-Audio-VAE along three axes: reconstruction fidelity on public benchmarks, encoding efficiency, and impact on downstream text-to-audio training. Reconstruction is measured on LibriSpeech~\citep{panayotov2015librispeech} for speech, AudioCaps~\citep{audiocaps} for general sound, and SongDescriber~\citep{songdescriber} for music, which together span the acoustic diversity the model is intended to cover. We compare against representative discrete neural codecs (AudioDec~\citep{audiodec}, DAC~\citep{dac}, EnCodec~\citep{Encodec}, and WavTokenizer~\citep{wavtokenizer}) and continuous audio VAEs (MM-Audio~\citep{mmaudio}, Stable Audio Open~\citep{stableaudioopen}, and Hunyuan-Foley~\citep{hunyuanfoley}).

We report Mel distance, multi-resolution STFT distance (MR-STFT), PESQ, and STOI. Since no single metric fully reflects perceptual quality---objective scores can, for example, overlook high-frequency muffling---we report all four and interpret them jointly: Mel distance and MR-STFT emphasize spectral fidelity, PESQ and STOI emphasize perceptual quality and intelligibility, and the latent frame rate reflects downstream modeling cost.

%
%
%
%
\subsection{Evaluation on Public Benchmarks}

The three benchmarks stress complementary abilities: LibriSpeech targets clean-speech intelligibility and the preservation of speaker identity and prosody; AudioCaps covers environmental sounds, sound effects, and mixed scenes; and SongDescriber emphasizes harmonic structure, timbre, and wide-band detail in music and singing. In each table (Tables~\ref{tab:librispeech-reconstruction}--\ref{tab:songdescriber-reconstruction}) we group models by latent frame rate, because a meaningful comparison must account for sequence length: Qwen-Audio-VAE competes directly with other low-frame-rate ($<$50\,Hz) systems, while high-frame-rate codecs are listed for reference.

\begin{table}[htbp]
\centering
\caption{Reconstruction results on LibriSpeech~\citep{panayotov2015librispeech}. Models are grouped by latent frame rate ($\geq$50\,Hz above the rule, $<$50\,Hz below); the best result within each group is in \textbf{bold}.}
\label{tab:librispeech-reconstruction}
\small
\begin{tabular}{lccccc}
\toprule
Model & Frame Rate & Mel Dist ↓ & MR-STFT ↓ & PESQ ↑ & STOI ↑ \\
\midrule
AudioDec & 80 Hz & 1.099 & 3.391 & 1.986 & 0.819 \\
DAC & 75 Hz & \textbf{0.305} & \textbf{0.784} & \textbf{4.464} & \textbf{0.995} \\
EnCodec & 75 Hz & 1.567 & 4.239 & 1.557 & 0.845 \\
WavTokenizer & 75 Hz & 1.104 & 3.437 & 2.380 & 0.912 \\
Hunyuan Foley & 50 Hz & 0.945 & 1.654 & 3.232 & 0.964 \\
\midrule
MM Audio & 31.25 Hz & 0.816 & 1.253 & 2.492 & 0.913 \\
Stable Audio Open & 20 Hz & 0.994 & 3.240 & 2.722 & 0.930 \\
\textbf{Qwen-Audio-VAE (Ours)} & 12.5 Hz & \textbf{0.513} & \textbf{0.715} & \textbf{3.975} & \textbf{0.980} \\
\bottomrule
\end{tabular}%
\end{table}

\begin{table}[htbp]
\centering
\caption{Reconstruction results on AudioCaps~\citep{audiocaps}. Models are grouped by latent frame rate ($\geq$50\,Hz above the rule, $<$50\,Hz below); the best result within each group is in \textbf{bold}.}
\label{tab:audiocaps-reconstruction}
\small
\begin{tabular}{lccccc}
\toprule
Model & Frame Rate & Mel Dist ↓ & MR-STFT ↓ & PESQ ↑ & STOI ↑ \\
\midrule
AudioDec & 80 Hz & 1.274 & 4.137 & 1.494 & 0.414 \\
DAC & 75 Hz & \textbf{0.320} & \textbf{0.958} & \textbf{4.392} & \textbf{0.974} \\
EnCodec & 75 Hz & 1.593 & 5.006 & 1.541 & 0.559 \\
WavTokenizer & 75 Hz & 1.374 & 3.990 & 1.477 & 0.475 \\
Hunyuan Foley & 50 Hz & 1.001 & 1.835 & 3.181 & 0.869 \\
\midrule
MM Audio & 31.25 Hz & 0.832 & 1.605 & 1.946 & 0.607 \\
Stable Audio Open & 20 Hz & 1.110 & \textbf{1.533} & 2.247 & 0.666 \\
\textbf{Qwen-Audio-VAE (Ours)} & 12.5 Hz & \textbf{0.649} & 2.315 & \textbf{2.952} & \textbf{0.812} \\
\bottomrule
\end{tabular}%
\end{table}

\subsubsection{LibriSpeech}

On LibriSpeech, Qwen-Audio-VAE is the best low-frame-rate model on every metric (Mel 0.513, MR-STFT 0.715, PESQ 3.975, STOI 0.980) while operating at only 12.5\,Hz. High-frame-rate codecs such as DAC reach higher raw fidelity, but only with $4$--$6\times$ longer latent sequences---precisely the cost we seek to avoid for downstream generation. We examine frame-rate and architecture variants in Section~\ref{sec:ablation}.

\subsubsection{AudioCaps}

On AudioCaps, whose diverse and transient-heavy content is harder than clean speech, Qwen-Audio-VAE again leads the low-frame-rate group on Mel distance, PESQ, and STOI, showing that its representation is not specialized to speech. Its MR-STFT is higher than Stable Audio Open's, a reminder that fine-grained spectral matching and perceptual intelligibility do not always move together on general audio.

\begin{table}[htbp]
\centering
\caption{Reconstruction results on SongDescriber~\citep{songdescriber}. Models are grouped by latent frame rate ($\geq$50\,Hz above the rule, $<$50\,Hz below); the best result within each group is in \textbf{bold}.}
\label{tab:songdescriber-reconstruction}
\small
\begin{tabular}{lccccc}
\toprule
Model & Frame Rate & Mel Dist ↓ & MR-STFT ↓ & PESQ ↑ & STOI ↑ \\
\midrule
AudioDec & 80 Hz & 1.206 & 2.631 & 1.162 & 0.472 \\
DAC & 75 Hz & 0.574 & 1.740 & 4.274 & 0.978 \\
EnCodec & 75 Hz & 1.110 & 2.389 & 1.391 & 0.634 \\
WavTokenizer & 75 Hz & 1.198 & 2.584 & 1.174 & 0.501 \\
Hunyuan Foley & 50 Hz & \textbf{0.503} & \textbf{1.033} & \textbf{4.400} & \textbf{0.987} \\
\midrule
Stable Audio Open & 20 Hz & 0.752 & 0.986 & 2.617 & 0.838 \\
\textbf{Qwen-Audio-VAE (Ours)} & 12.5 Hz & \textbf{0.671} & \textbf{0.903} & \textbf{3.319} & \textbf{0.883} \\
\bottomrule
\end{tabular}%
\end{table}

\begin{figure}[htbp]
\centering
\includegraphics[width=\textwidth]{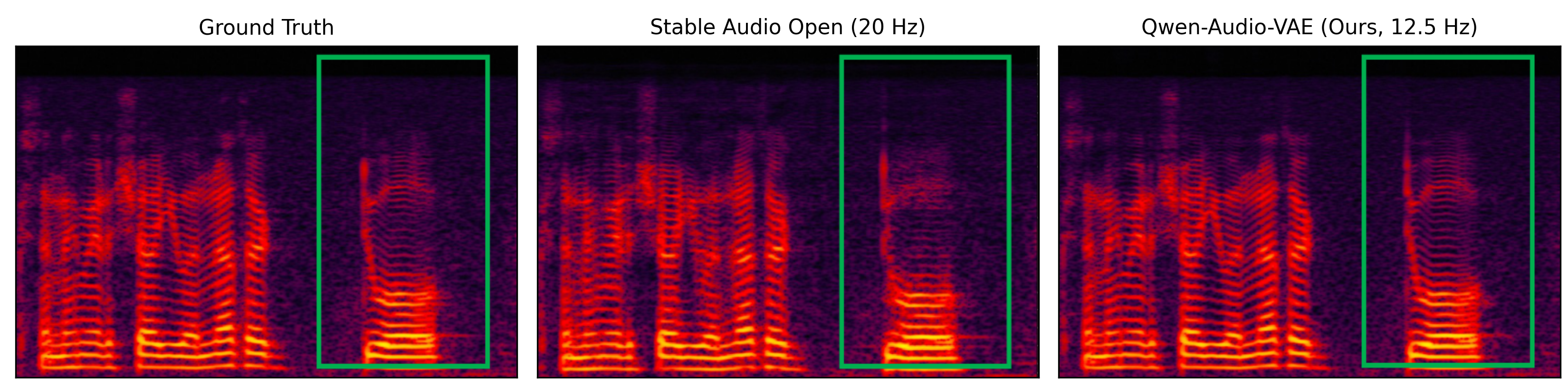}
\caption{Spectrogram reconstruction comparison on LibriSpeech. Columns: ground truth, Stable Audio Open (20\,Hz), and Qwen-Audio-VAE (Ours, 12.5\,Hz); the green box marks the highlighted high-frequency region. Despite its lower latent frame rate, Qwen-Audio-VAE preserves harmonic structure and high-frequency texture close to the ground truth, whereas Stable Audio Open blurs high-frequency detail.}
\label{fig:spectrogram-compare}
\end{figure}

\subsubsection{SongDescriber}

On SongDescriber, Qwen-Audio-VAE surpasses Stable Audio Open on all four metrics, showing that its compact latent also captures musical harmonic and timbral structure; among high-frame-rate systems the music-oriented Hunyuan-Foley is strongest, as expected for a 50\,Hz model tuned for music.

Taken together, the three benchmarks establish the intended operating point of Qwen-Audio-VAE: it is consistently the best low-frame-rate model across speech, sound, and music while using the fewest latent frames, trading a small margin of raw fidelity against high-frame-rate codecs for a $4$--$6\times$ shorter latent sequence. Figure~\ref{fig:spectrogram-compare} illustrates this qualitatively on LibriSpeech: our 12.5\,Hz reconstruction tracks the ground-truth spectrum more faithfully than the higher-frame-rate Stable Audio Open, most visibly in the highlighted high-frequency region.

\subsection{Ablation Study}
\label{sec:ablation}

We isolate individual design choices in Table~\ref{tab:ablation}, using LibriSpeech as a representative benchmark; all rows share the encoder/decoder topology of the main model and differ only in the factor being ablated.

\begin{table}[htbp]
\centering
\caption{Ablation of Qwen-Audio-VAE variants, evaluated on LibriSpeech~\citep{panayotov2015librispeech}. Variants differ in latent frame rate, the encoder-side window Transformer, latent dimension, and decoder width.}
\label{tab:ablation}
\small
\begin{tabular}{lccccc}
\toprule
Variant & Frame Rate & Mel Dist ↓ & MR-STFT ↓ & PESQ ↑ & STOI ↑ \\
\midrule
Qwen-Audio-VAE (main) & 12.5 Hz & 0.513 & 0.715 & 3.975 & 0.980 \\
\midrule
Qwen-Audio-VAE-50Hz & 50 Hz & \textbf{0.467} & \textbf{0.637} & \textbf{4.040} & \textbf{0.987} \\
w/o Encoder Transformer & 12.5 Hz & 0.620 & 0.728 & 3.711 & 0.973 \\
64 dim & 12.5 Hz & 0.717 & 0.807 & 3.129 & 0.951 \\
64 dim + big decoder & 12.5 Hz & 0.715 & 0.801 & 3.155 & 0.953 \\
\bottomrule
\end{tabular}%
\end{table}

\paragraph{Encoder Transformer.} Removing the bottleneck window Transformer raises Mel distance from 0.513 to 0.620 and lowers every other metric, confirming that attention-based aggregation is valuable under aggressive temporal compression, where each latent frame must summarize a long audio span.

\paragraph{Latent dimension.} Halving the latent to 64 dimensions consistently hurts reconstruction, and widening the decoder does not recover it. The limiting factor is therefore how information is organized in the latent, not decoder capacity.

\paragraph{Frame rate.} The 50\,Hz variant is best on every metric, but at four times the sequence length. Because downstream text-to-audio cost grows with sequence length, the 12.5\,Hz model offers the better efficiency--quality trade-off and is our default configuration.

%
%
%
%
%
%
\subsection{Encoding Efficiency}

We now turn from reconstruction quality to the efficiency axis. Table~\ref{tab:encoding-efficiency-en} reports the wall-clock latency of encoding 64 clips of 30\,s, the representative workload of offline latent extraction for text-to-audio training. The three optimizations of Section~\ref{sec:encoder-accel} reduce latency from 1957\,ms to 541\,ms ($3.62\times$), with most of the gain coming from shifting computation away from the high-resolution early stages while preserving the capacity of the later, low-resolution ones.

\begin{table}[htbp]
\centering
\caption{Encoding efficiency comparison. $^{\dagger}$Stable Audio Open is 1.19$\times$ faster than Qwen-Audio-VAE \emph{before} acceleration, but remains slower than the optimized model (541\,ms).}
\label{tab:encoding-efficiency-en}
\small
\begin{tabular}{lcc}
\toprule
Model / Variant & 64$\times$30s Encoding Latency ↓ & Relative Speedup \\
\midrule
Qwen-Audio-VAE before acceleration & 1957 ms & 1.00$\times$ \\
+ Encoder stride optimization & 1361 ms & 1.44$\times$ \\
+ Residual unit pruning & 747 ms & 2.62$\times$ \\
+ First-layer channel reduction & \textbf{541 ms} & \textbf{3.62$\times$} \\
Stable Audio Open$^{\dagger}$ & 1640 ms & 1.19$\times$ \\
\bottomrule
\end{tabular}%
\end{table}

Low frame rate and fast encoding are complementary: the former shortens downstream DiT sequences, while the latter lowers the cost of producing the latents themselves. Notably, Stable Audio Open is slower than our optimized encoder despite running at a higher 20\,Hz frame rate, confirming that the two properties are independent advantages rather than a single speed-versus-rate trade-off.

\subsection{Text-to-Audio Integration}

We first quantify how encoder acceleration affects text-to-audio training. We train the same 150M-parameter diffusion transformer on latents produced by our encoder \emph{before} and \emph{after} acceleration, on 32$\times$L20 (80\,GB) GPUs with \textbf{LAION-Audio-630K}~\citep{laionclap2023} (4{,}325.39\,h after filtering). At a fixed throughput of $\sim$1 step/s, removing the encoding bottleneck lets the accelerated model use a $4\times$ larger batch (64 vs.\ 16) and process $4\times$ more audio per step, which lifts AudioCaps CLAP from 0.29 to 0.33 under the same wall-clock budget (Table~\ref{tab:text2audio-integration-en}). For reference, we also run a Stable Audio Open baseline through the same pipeline; its higher-frame-rate, slower encoder needs longer training yet reaches lower quality (60\,h for CLAP 0.27, versus 36\,h for our 0.33).

\begin{table}[htbp]
\centering
\caption{Text-to-audio training efficiency and generation quality. CLAP is measured on the AudioCaps test set.}
\label{tab:text2audio-integration-en}
\small
\setlength{\tabcolsep}{5pt}
\begin{tabular}{lccccc}
\toprule
Setting & GPU Setup & \makecell{Batch\\Size} & \makecell{Audio\\per Batch} & \makecell{Training\\Time} & \makecell{AudioCaps\\CLAP $\uparrow$} \\
\midrule
Stable Audio Open & 32$\times$L20 80G & 16 & 160\,s & 60\,h & 0.27 \\
Ours w/o Encoder Acceleration & 32$\times$L20 80G & 16 & 160\,s & 36\,h & 0.29 \\
Ours & 32$\times$L20 80G & \textbf{64} & \textbf{640\,s} & \textbf{36\,h} & \textbf{0.33} \\
\bottomrule
\end{tabular}
\end{table}

Beyond throughput, the structure of the latent space itself shapes downstream generation. Starting from our default latent (KL weight $\lambda_{\text{kl}}=10^{-6}$), we trained a more strongly regularized variant ($\lambda_{\text{kl}}=10^{-3}$), selecting a checkpoint that balances reconstruction and KL loss, and evaluated both with FD (VGG), Inception Score (IS), and CLAP on the AudioCaps test set (Table~\ref{tab:latent-kl}). The more strongly regularized latent \emph{degrades} generation on all three metrics. We attribute this to latent information density: pulling the posterior harder toward the prior discards structure the transformer relies on, and we separately observe that higher-dimensional latents are harder for the diffusion transformer to fit. Aligning the latent with pretrained audio representations (e.g., CLAP or an Omni audio encoder) is a promising direction for improving generative learnability.

\begin{table}[htbp]
\centering
\caption{Effect of stronger KL regularization on text-to-audio generation (AudioCaps test set; the same 150M diffusion transformer). Raising the KL weight from $10^{-6}$ to $10^{-3}$ regularizes the latent but hurts every generation metric.}
\label{tab:latent-kl}
\small
\begin{tabular}{lccc}
\toprule
Setting & FD (VGG) $\downarrow$ & IS $\uparrow$ & CLAP $\uparrow$ \\
\midrule
\textbf{Ours ($\lambda_{\text{kl}}=10^{-6}$)} & \textbf{2.44} & \textbf{8.94} & \textbf{0.33} \\
Ours ($\lambda_{\text{kl}}=10^{-3}$) & 5.18 & 7.09 & 0.30 \\
\bottomrule
\end{tabular}
\end{table}

\section{Conclusion}
\label{sec:conclusion}

We presented Qwen-Audio-VAE, a suite of continuous audio autoencoders for scalable general audio generation. Its design is organized around three goals that a generation backbone must meet simultaneously: high-fidelity reconstruction across diverse audio, a compact latent that keeps downstream sequences short, and high-throughput encoding for large-scale training. The model couples a causal encoder--decoder with a continuous 12.5\,Hz bottleneck, concentrates window Transformers at that low rate, and adds a bank of complementary discriminators for adversarial supervision. Trained on 5 million hours of multi-domain audio, it reconstructs speech, music, and general sound faithfully across heterogeneous acoustic conditions.

Across public benchmarks (LibriSpeech, AudioCaps, SongDescriber), Qwen-Audio-VAE is consistently the strongest low-frame-rate model while using the fewest latent frames. Its asymmetric encoder--decoder and latency-aware encoder pruning cut 64$\times$30s encoding latency from \textbf{1957\,ms to 541\,ms} ($3.62\times$) without degrading reconstruction, and this speedup translates into system-level gains: in text-to-audio training, faster latent extraction supports a $4\times$ larger batch and higher throughput at the same training time. More broadly, our results suggest that autoencoders for general audio generation should be optimized not for reconstruction alone, but jointly for compression, encoding throughput, data scalability, and latent learnability. Future work will pursue aligning the latent with pretrained audio representations, improving high-frequency and transient reconstruction, strengthening data stratification, and validating on larger-scale text-to-audio training.

%
%
%

\clearpage
\bibliography{biblio}
\bibliographystyle{colm2024_conference}

\clearpage
\section{Core Contributors}

\vspace{2pt}
{\fontsize{9}{10}\selectfont
\renewcommand{\thefootnote}{\fnsymbol{footnote}}%
\begin{tabular}{@{}*{6}{l@{\hspace{1.4em}}}@{}}
Ziyue Jiang & Dake Guo & Zekai Zhang & Hangrui Hu & Ting He & Xinfa Zhu \\
Xiong Wang & Yongqi Wang & Jiapeng Wang & Wenxiang Guo & Zhifang Guo & Chenfei Wu \\
Dayiheng Liu & Jin Xu\footnotemark[2]\footnotemark[1] & & & & \\
\end{tabular}
\footnotetext[2]{Team Lead.}
\footnotetext[1]{Corresponding Author.}
\par}

\clearpage
\section{Appendix}
\label{sec:appendix}

\subsection{Model Configuration}
\label{sec:appendix-config}

For completeness and reproducibility, Table~\ref{tab:model-config} reports the full configuration of the main Qwen-Audio-VAE model in signal-flow order; its key values also appear inline in Section~\ref{sec:components}.

\begin{table}[htbp]
\centering
\caption{Main Qwen-Audio-VAE configuration (24\,kHz, mono). Ablated variants are described in Section~\ref{sec:experiment} and Appendix~\ref{sec:appendix-variants}.}
\label{tab:model-config}
\small
\begin{tabularx}{\textwidth}{@{}llX@{}}
\toprule
Module & Hyperparameter & Value \\
\midrule
\multirow{2}{*}{I/O}
 & Audio & 24\,kHz, mono \\
 & Latent & 12.5\,Hz, 128-d, diagonal Gaussian \\
\midrule
\multirow{3}{*}{Encoder (causal)}
 & Downsampling & strides $8{,}5{,}4{,}3$ ($480\times$, $\to$ 50\,Hz) \\
 & Channels & $24\!\to\!128\!\to\!256\!\to\!512\!\to\!1024$ \\
 & Residual units / block & 1 (dilation $1$) \\
\midrule
\multirow{2}{*}{Bottleneck}
 & Extra downsampling & $4\times$ (50\,Hz $\to$ 12.5\,Hz) \\
 & Window Transformer (pre \& post) & 8 layers, dim 1024, 16 heads, window 72 \\
\midrule
\multirow{3}{*}{Decoder (causal)}
 & Upsampling & strides $8{,}5{,}4{,}3$ ($480\times$, $\to$ 24\,kHz) \\
 & Base channels & 1536 \\
 & Residual units / block & 3 (dilations $1{,}3{,}9$) \\
\midrule
Discriminators & Types & multi-period, multi-resolution STFT, multi-scale STFT, sub-band CQT \\
\bottomrule
\end{tabularx}
\end{table}

\subsection{Training Details}
\label{sec:appendix-training}

The model is trained from scratch on the full multi-domain corpus. Table~\ref{tab:training-config} lists the complete setup, including the optimizer, the separate generator/discriminator learning rates, the learning-rate schedule, and the hardware.

\begin{table}[htbp]
\centering
\caption{Training configuration of Qwen-Audio-VAE.}
\label{tab:training-config}
\small
\begin{tabular}{ll}
\toprule
Hyperparameter & Value \\
\midrule
Training data & 5M hours, multi-domain (speech / music / sound) \\
Sample rate & 24\,kHz (mono) \\
Training segment length & 6\,s (144{,}000 samples) \\
Optimizer & AdamW, $\beta=(0.5, 0.9)$, $\epsilon=10^{-9}$ \\
Learning rate (G / D) & $2\times10^{-4}$ / $1\times10^{-4}$ \\
LR schedule & linear warmup 8{,}000 steps; decay from step 400{,}000 \\
Batch size & 2 per GPU, gradient accumulation 1 \\
Total steps & 2.4M \\
Gradient clipping & max-norm 1.0 \\
GPUs & 32$\times$A100 80G \\
Latent frame rate / dim & 12.5\,Hz / 128 \\
\bottomrule
\end{tabular}
\end{table}

\subsection{Discriminator Configuration}
\label{sec:appendix-disc}

Section~\ref{sec:components} introduced the four discriminators used for adversarial training; Table~\ref{tab:disc-config} gives their exact periods, FFT sizes, and constant-$Q$ resolutions.

\begin{table}[htbp]
\centering
\caption{Discriminator configuration.}
\label{tab:disc-config}
\small
\begin{tabularx}{\textwidth}{@{}lX@{}}
\toprule
Discriminator & Configuration \\
\midrule
Multi-period~\citep{hifigan} & periods $= \{2,3,5,7,11\}$ \\
Multi-resolution STFT~\citep{dac} & FFT sizes $= \{334,542,876,1418,2296\}$; 5 sub-bands \\
Multi-scale STFT~\citep{Encodec} & filters $=72$; $n_\text{FFT} = \{206,334,542,876,1418,2296\}$ \\
Sub-band CQT & filters $=128$; dilations $\{1,2,4\}$; octaves $\{9,9,9\}$; bins/oct. $\{24,36,48\}$ \\
\bottomrule
\end{tabularx}
\end{table}

\subsection{Model Variants}
\label{sec:appendix-variants}

Table~\ref{tab:variants} maps the ablation rows reported in Section~\ref{sec:ablation} to the configuration fields that distinguish them. All variants share the main model's 24\,kHz encoder/decoder topology (strides $8,5,4,3$) and differ only in the latent dimension (set by the bottleneck projection), the encoder-side window Transformer, the decoder width, and the bottleneck downsampling factor that sets the final frame rate.

\begin{table}[htbp]
\centering
\caption{Configuration of model variants.}
\label{tab:variants}
\small
\begin{tabularx}{\textwidth}{@{}lcccX@{}}
\toprule
Variant & Frame rate & Latent dim & Decoder dim & Note \\
\midrule
\textbf{Qwen-Audio-VAE (main)} & 12.5\,Hz & 128 & 1536 & full model \\
w/o Encoder Transformer & 12.5\,Hz & 128 & 1536 & encoder-side window Transformer removed \\
64 dim & 12.5\,Hz & 64 & 1536 & smaller latent \\
64 dim + big decoder & 12.5\,Hz & 64 & 2048 & smaller latent + wider decoder \\
50\,Hz variant & 50\,Hz & 128 & 1536 & no extra bottleneck downsampling \\
\bottomrule
\end{tabularx}
\end{table}

\end{document}